\documentclass[twoside,12pt]{article}
\usepackage{amsmath} 
\usepackage{CJK}   
\usepackage{indentfirst}  
\usepackage{bm}    
\usepackage{graphicx}  
\usepackage{cases}
\usepackage[super,square,comma,sort&compress]{natbib}
\usepackage{graphicx}
\usepackage{float}
\usepackage{float,amsmath, amssymb,epsfig,graphicx,mathrsfs}
\usepackage{amsfonts,bbm,color,bm}
\usepackage{amsmath}

\begin{document}    

\begin{CJK*}{GBK}{song}  



\begin{center}
\LARGE\bf Satellite-based continuous-variable quantum key distribution under the Earth's gravitational field$^{*}$   
\end{center}

\footnotetext{\hspace*{-.45cm}\footnotesize $^\dag$ Corresponding author. E-mail: caoshuo@bnu.edu.cn; xiaobaoliu@hotmail.com;}

\begin{center}
\rm Tonghua Liu $^{1,2}$,  Shuo Cao$^{2,3\rm \dag}$, Sixuan Zhang$^{4}$, Shuai Ma$^{5}$, and Xiaobao Liu$^{6\rm \dag}$
\end{center}

\begin{center}
\begin{footnotesize} \sl
${}^{\rm} $1. School of Physics and Optoelectronic, Yangtze University, Jingzhou 434023, China; \\
 2. Frontier Science Institute of Astronomy and Astrophysics, Beijing Normal University, Beijing, 100875, China; \\
 3. Department of Astronomy, Beijing Normal University, Beijing 100875, China;\\
 4. Graduate School of Advanced Science and Engineering, Hiroshima University, Hiroshima 739-8526, Japan;\\
 5. Beijing Academy, Kangyuan Road, Dongba, Beijing 100018, China; \\
 6. Department of physics and electrical engineering, Liupanshui Normal University, Liupanshui 553004, Guizhou, China.
\end{footnotesize}
\end{center}

\begin{center}
\footnotesize (Received X XX XXXX; revised manuscript received X XX XXXX)
\end{center}

\vspace*{1mm}

\begin{center}
\begin{minipage}{15.5cm}
\parindent 20pt\footnotesize

Long distance communication protocols cannot ignore the existence of the Earth's gravitational field and its effects on quantum states. In this work, we show a very general method to consider the effects of the Earth's gravitational field on continuous-variable quantum key distribution protocols. Our results show that the Earth's gravitational field erodes the ability of the two parties to perform QKD in all the protocols. However, our findings also exhibit some interesting features, i.e., the key rates initially increase for a specific range of height parameter $h\simeq r_A/2$ and then gradually decrease with the increasing of the orbits of satellite $h$. A possible explanation is also provided in our analysis, considering the fact that gravitational frequency shift and special relativistic effects play different roles in the key rates. In addition,  our findings show that the change in key rate effected by gravitational frequency shift can be determined at a level of $<1.0\%$ within the satellite height at geostationary Earth orbits.   Our work could provide some interesting possibilities to reduce the loss key rate through the control of the orbital height of satellites.

\end{minipage}
\end{center}

\begin{center}
\begin{minipage}{15.5cm}
\begin{minipage}[t]{2.3cm}{\bf Keywords:}\end{minipage}
\begin{minipage}[t]{30.1cm}
Quantum information; Quantum key distribution; Relativistic effect.

\end{minipage}\par\vglue8pt
{\bf PACS: 04.62.+v, 03.65.Ud, 06.20.-f, 03.67.-a}
\end{minipage}
\end{center}

\section{introduction}

The most important part of quantum communication is quantum key distribution (QKD). In simple terms, QKD is a secure key transmission method that can be used to send keys between two remote communication ends.  In the process of secure communication, information needs to be encrypted and decrypted with a key. The security of the key ensures the security of information, which belongs to solving the key problem. The security of QKD is guaranteed by the quantum measurement theory and the quantum no cloning theorem. Therefore, any eavesdropping on the QKD process could change the quantum states themselves, resulting in high bit error rates that allow eavesdropping to be detected. Generally speaking, the transmission of quantum state in the QKD process relies on the encoding, transmitting and measuring of photons. In 1984, \textit{Bennett} and \textit{Brassart} proposed the BB84 protocol \cite{1}, which used quantum bits as the information carrier and used the polarization characteristics of light to encode the quantum state, so as to realize the generation and secure distribution of the secret key. Subsequently, \textit{Ekert} reported the practical application of the generalized Bell's theorem in the key distribution process in cryptography \cite{1a}. This kind of protocol is called discrete-variable quantum key distribution (DV-QKD) protocols. In the QKD process, the quantum correlations of entangled photons are usually used to generate secure cryptographic keys. Hence, a key challenge in this context is to establish  continuously working, reliable long-distance entanglement distributions. Fortunately, the quantum information task has been a great success experimentally, such as the experiment in refs \cite{2b,2c}. Such as \textit{Yin et al.} demonstrated the entanglement distribution of two photons from a satellite to two ground stations over long distances, which illustrates the possibility of a future global quantum communication network \cite{3,4}.  Especially, there are a number of experiments to perform QKD protocols \cite{5,6,7}, which include measurement-device-independent \cite{8} and single-photon interference version twin-field QKD \cite{9} and so on. For more related work on QKD, we refer readers to the references \cite{9a,9b,9c,21a}.

On the other hand, QKD protocols provide the ultimate security against channel attacks, but their practical implementations are still challenging. Many methods require trusted experimental devices and detectors and are therefore subject to so-called side-channel attacks against these devices.
Measurement device independent (MDI) is an important type of QKD protocols among them \cite{8,10b}. MDI QKD provides a middle ground, relaxing the assumptions on the protocol by having distant parties send states to a central relay detector which may be controlled by an Eavesdropper (Eve). Moreover, point-to-point quantum communications are known to be inherently distance limited by the PLOB bound \cite{10q} with the transmissivity decaying exponentially with distance. Continuous variable (CV) QKD protocols \cite{2,2a,10c,10d,10f,10e} are able to reach rates approaching the PLOB bound and have a longer transmission distance, outperforming discrete state protocols \cite{10g,10h}.  More relevant reviews about CV QKD and Gaussian quantum information we suggest readers to see refs \cite{10,11,CVQKB1,CVQKB2,CVQKB3,CVQKB4,CVQKB5,CVQKB6,CVQKB7,CVQKB8,CVQKB9,CVQKB10}. In addition, an asynchronous measurement-device-independent quantum key distribution protocol has been proposed recently \cite{CVQKB11,CVQKB12}. These work have improved the secret key capacity and broke the rate-distance limit by postmatching two interference detection events to realize asynchronous two-photon Bell-state measurement.
In general, the standard protocols in CV QKD consider that Alice prepares squeezed states with a Gaussian modulation along one of the two quadratures initially, and sends them to Bob, who measures them by performing homodyne detection. However, long-distance quantum information tasks and protocols usually require the assistance of satellites which allows us to extend techniques, such as QKD, to global scales. In this context, the effect of the gravity of the Earth should be considered at long distance quantum communication. Therefore, we consider CV QKD scenario in the analysis, due to CV MDI QKD protocols further boast of longer transmission distances in comparison to the discrete variable counterparts.

Since the quantum system in reality always shows the characteristics of gravity and relativity, the quantum system cannot be prepared and transmitted without the existence of a gravitational field. When quantum resources are studied in a relativistic setting, the effects of gravity and accelerated motion, especially on the quantum properties and their applications, have always been ignored, which fails to overcome the inherent inconsistency between quantum physics and relativity. Fortunately, the quantum field theory of curved space-time provides a theoretical framework to carry out the analysis above. The effect of gravity can be considered as a curved dynamic background, thus correcting the equations of quantum field theory. In fact, it has been clarified only recently how Gaussian quantum states are properly affected by gravitational field or acceleration. For instance, the work has investigated how relativistic acceleration affects the performance of quantum teleportation and dense coding for CV states of localized wave packets \cite{12}. More works about relativistic acceleration or the Earth's gravity affect on quantum resources (entanglement, coherence and Gaussian Interference Power) and quantum information tasks, we refer readers to see refs \cite{13,14,wu1,wu2,wu3,15,16,17,18,19,20,21}. From an experimental point of view, \textit{Pirandola et al.} proposed a fully operational satellite QKD system \cite{9b}, which means that their scheme paves the way toward the implementation of a quantum communication worldwide network leveraging existing receivers.  Furthermore, from a theoretical point of view, \textit{Pierini} provided a general method to investigate how noninertial motion affects the performance of various QKD protocols for continuous-variable localized Gaussian states and stated that his results can be applied to the gravitational field according to the equivalence principle \cite{22}.

Inspired by the above, we will focus on the QKD of CV system under the gravitational effect of the Earth when one part of a bipartite system is involved in relativistic effects. We will quantitatively analyze how the curved space-time background of the Earth affects QKD and the key rate. We assume that the entangled photon pair is initially prepared in a two-mode compressed state, and then assume that one of the photon pairs stays on the Earth's surface while the other photon pair propagates to the satellite. Due to the Earth's gravitational field, the wave-packet of photon will be deformed in the propagating process when sending it to the satellite.  Therefore,  we adopt a lossy quantum channel to describe this deformed effect on the quantum state of photons.  This paper is organized as follows:  In Sec. 2, we introduce the general protocol for QKD with CV and  key rates. In Sec. 3, we briefly describe the propagation of a photon wave packet from the Earth to a satellite. The influence of the Earth's gravitational field on the previously described protocol for QKD, in particular on the efficiency in distributing the secret key is presented in Sec. 4.  Our results and discussions are given in Sec. 5.  Finally, we summarize our conclusions in Sec. 6. The natural units $G = c = \hbar = k_B =  1$ are employed throughout the paper.

\section{Continuous-variable quantum key distribution and secret key}
A standard QKD protocol of CV can be briefly described as follows: firstly, Alice prepares a two-mode squeezed vacuum state and one of them is sent to Bob. Then Alice and Bob randomly measure either $x$ or $p$ quadrature of their shared entangled state. Among all the data collected after measurements, they keep only the outcomes of the measurements done on the same basis. This procedure is called sifting. After that, they perform the sifting procedure, discarding data that of the result taken from different bases measurements. Eve wants to learn the key by using an interaction with the quantum system. This action, in general, modifies the state in a way that can be observed.  These perturbations allow Alice and Bob to infer that some eavesdropping is happening and, consequently, estimate the amount of information leaked to Eve.
Thus, in the second step, the reliable parts reveal randomly selected data samples to each other to estimate the parameters of the channel and thus estimate the amount of information leaked to Eve. After sifting and parameter estimation, Alice and Bob share a string of related elements, called the original key. In the third step, they extract a common binary key from their data using classical communication.  Depending on whether Alice's or Bob's data is used, the protocol is called direct reconciliation (DR) or reverse reconciliation (RR). Finally, the perfect correlation key is obtained through error correction and privacy amplification.

In the standard QKD procedure,  we assume that Eve is passively attacked,  she replaces the real connection between Alice and Bob with a unitary operation that mimics the channel between them when we trace Eve's modes. This operation is represented by a thermal noise channel of transmissivity $T$ and noise referred to the input $\chi=\frac{1-T}{T}+\epsilon$, where $\epsilon$ is the so-called excess noise.  In
practice, we can use a beam splitter of parameter $T$ for implementation, where the signal sent to Bob is combined with
a thermal state of zero mean and variance $\langle\hat{X}_{th}^{2}\rangle=\frac{T}{1-T}\chi$. The process can be described by the following form
\begin{equation}
\hat{X}_B=\sqrt{T}\hat{X}_A+\sqrt{1-T}\hat{X}_{th}.
\end{equation}
This attack is called an intercept and resend or clone attack. To determine whether the set of common binary symbols created in the protocol can be used for encryption tasks, the two honest parts must evaluate the secret key rate $K$. In general, Eve has two ways to perform attacks, i.e., individual attacks (IA) and collective attacks (CA), which result in two kinds of secret key rate. For convenience, we only consider the CA and the DR in this paper when analyzing the secret key rate.

The secret key rate of QKD protocol in CA and the DR case is calculated as \cite{22,23}
\begin{equation}
K=I(a:b)-S(a:E)
\end{equation}
when Eve performs an optimal collective measurement. Here, the $I(a:b)$ is the mutual information between Alice and Bob, and the $S(a:E)$ is the Holevo bound on Eve information, the eavesdropper, stolen when considering reverse reconciliation. The first term can be calculated by
\begin{equation}
I(a:b)=\frac{1}{2}\log_2\frac{V_A}{V_{A|B}},
\end{equation}
where $V_A$ is the modulation variance and $V_{A|B}$ is the conditional variance of the random variable of Alice while knowing Bob's measurement results. For second term in secret key rate $K$,  the Holevo bound for DR can be written as
\begin{eqnarray}
S(a:E)=S(E)-S(E|a).
\end{eqnarray}
Assuming that Eve holds the purification of the state allows us to write $S(E) = S(AB)$ and $S(E|a) = S(B|a)$, so that Eve's accessible information is only a function of the entropic quantities of Alice and Bob. For a given two modes covariance matrix
\begin{equation}
\sigma_{AB}=
\left(
\begin{array}{cc}
A& C\\
C^T&B
\end{array}
\right),
\end{equation}
where $A$ and $B$ are $2\times2$ covariance matrices of the respective subsystems and $C$ is the correlation matrix, we have that
\begin{eqnarray}
S(AB)=g(\lambda_1)+g(\lambda_2),
\end{eqnarray}
where $\lambda_{1,2}$ are the symplectic eigenvalues of
\begin{eqnarray}
\lambda_{1,2}&=&\sqrt{\frac{1}{2}(\triangle\pm\sqrt{\triangle^2-4D^2})},\nonumber\\
\triangle&=& \mathrm{det} A+ \mathrm{det} B+2 \mathrm{det} C,\nonumber\\
D&=&\mathrm{det} \sigma_{AB},
\end{eqnarray}
and
\begin{eqnarray}
g(x)=(\frac{x+1}{2})\log_2(\frac{x+1}{2})-(\frac{x-1}{2})\log_2(\frac{x-1}{2}).
\end{eqnarray}
Also, we have that
\begin{eqnarray}
S(B|a)=g(\lambda_3),
\end{eqnarray}
$\lambda_3=\sqrt{\mathrm{det} (\sigma_{B|a})}$ being the symplectic eigenvalue of Bob's covariance matrix $\sigma_{B|a}$ given Alice measurement outcome. When Alice performs measurement on his
mode, which gives $a$ as result, the covariance matrix
\begin{eqnarray}
\sigma_{B|a}=B-C(A_{11}^{-1}\Pi)C^T,
\end{eqnarray}
where $\Pi=\left(\begin{array}{cc}1& 0\\0&0\end{array}\right)$. This is a squeezed state displaced along the $x$ quadrature.

\section{The propagation of photons from the Earth to satellite}
In this section, we will describe the propagation of photons from the Earth to satellites. In this context, photons will suffer the effects of the Earth's gravitational field.  It is well known that Earth is a spherical planet with angular velocity $\omega=7.29\times10^{-5}$ rad/s. Thus, we adopt the  Kerr metric to describe the Earth's space-time approximately. For the convenience of calculation, we restricted our work to the equatorial plane $\theta=\frac{\pi}{2}$. The reduced metric in Boyer-Lindquist coordinates $(t,r,\phi)$ has following form  \cite{Visser,kerr}
\begin{align}\label{metric}
ds^2=&\, -\Big(1-\frac{2M}{r} \Big)dt^2+\frac{1}{\Delta}dr^2 \nonumber \\
&\,+\Big(r^2+a^2+\frac{2Ma^2}{r}\Big) d\phi^2 - \frac{4Ma}{r} dt \, d\phi, \\
\Delta=&\,1-\frac{2M}{r}+\frac{a^2}{r^2},
\end{align}
where $M$, $r$, $J$, $a=\frac{J}{M}$ are mass, radius, angular momentum and Kerr parameter of the Earth, respectively.

Assuming that Alice on Earth's surface prepares a monochromatic photon and sends it to Bob at time $\tau_A$, Bob will receive this photon at $\tau_B=\Delta\tau+\sqrt{f(r_B)/f(r_A)}\tau_A$ in his own reference frame, where $f(r_A)=1-\frac{r_S}{r_A}$ and  $f(r_B)=1-\frac{r_S}{r_B}$, $r_S=2M$ is the Schwarzschild radius of
 Earth and $\Delta\tau$ is the propagation time of the light from Earth to satellites by taking the curved effects of the Earth into account. In general, a photon can be modeled by a wave packet of massless bosonic fields with a distribution $F^{(K)}_{\Omega_{K,0}}$ of mode frequency $\Omega_{K}$ and peaked at $\Omega_{K,0}$ \cite{ULMQ,TGDT}, where $K=A,B$ denotes the modes in Alice or Bob reference frames, respectively. The annihilation operator of a photon for an observer far from Alice or Bob given in refs \cite{DEBT,DEBA}, which their works proposed a quantum experiment to measure  Schwarzschild spacetime parameters of the Earth with high precision
\begin{equation}
a_{\Omega_{K,0}}(t_K)=\int_0^{+\infty}d\Omega_K e^{-i\Omega_K t_K}F^{(K)}_{\Omega_{K,0}}(\Omega_K)a_{\Omega_K}.
\label{wave}
\end{equation}
Alice's and Bob's operators in Eq. (\ref{wave}) can be used to describe the same optical mode at different altitudes. By considering the spacetime background of the Earth,
 the wave packet received is modified. The relation between $a_{\Omega_A}$ and $a_{\Omega_B}$ was discussed in refs \cite{15,16,17}, and can be used to obtain the relation between the frequency distributions $F^{(K)}_{\Omega_{K,0}}$ of the photons before and after the propagation
\begin{eqnarray}
F^{(B)}_{\Omega_{B,0}}(\Omega_B)=\sqrt[4]{\frac{f(r_B)}{f(r_A)}}F^{(A)}_{\Omega_{A,0}}\left(\sqrt{\frac{f(r_B)}{f(r_A)}}\Omega_B\right).\label{wave:packet:relation}
\label{fab}
\end{eqnarray}
In Eq. (\ref{fab}), we can see that the effects induced by the
curved spacetime background of the Earth cannot be simply corrected by a linear shift of frequencies.

Fortunately, following such a nonlinear gravitational effect, one can use the fidelity of the quantum channel to model this effect. It is always possible to decompose the mode $\bar{a}^{\prime}$ received by Bob in terms of the mode $a^{\prime}$ prepared by Alice and an orthogonal mode $a_{\bot}^{\prime}$ (i.e. $[a^{\prime},a_{\bot}^{\prime\dagger}]=0$) \cite{PPRW}
\begin{eqnarray}
\bar{a}^{\prime}=\Theta a^{\prime}+\sqrt{1-\Theta^2}a_{\bot}^{\prime},\label{mode:decomposition}
\end{eqnarray}
where $\Theta$ is the wave packet overlap between the distributions $F^{(B)}_{\Omega_{B,0}}(\Omega_B)$ and $F^{(A)}_{\Omega_{A,0}}(\Omega_B)$
\begin{eqnarray}
\Theta:=\int_0^{+\infty}d\Omega_B\,F^{(B)\star}_{\Omega_{B,0}}(\Omega_B)F^{(A)}_{\Omega_{A,0}}(\Omega_B),\label{single:photon:fidelity}
\end{eqnarray}
and we have $\Theta=1$ for a perfect channel. If the curvature is strong enough, the fidelity $\mathcal{F}=|\Theta|^2$ would be low.

We assume that Alice employ a real normalized Gaussian wave packet of the form
\begin{eqnarray}
F_{\Omega_0}(\Omega)=\frac{1}{\sqrt[4]{2\pi\sigma^2}}e^{-\frac{(\Omega-\Omega_0)^2}{4\sigma^2}}\label{Bobpacket},
\end{eqnarray}
with wave packet width $\sigma$. In this case, the overlap $\Theta$ is given by \eqref{single:photon:fidelity} where we have extended the domain of integration to all the real axis. We note that the integral should be performed over strictly positive frequencies. However, since $\Omega_0\gg \sigma$, it is possible
to include negative frequencies without affecting the value of $\Theta$. Using Eqs. \eqref{wave} and \eqref{Bobpacket} one can find that
\begin{eqnarray} \label{theta}
\Theta=\sqrt{\frac{2}{1+(1+\delta)^2}}\frac{1}{1+\delta}e^{-\frac{\delta^2\Omega_{B,0}^2}{4(1+(1+\delta)^2)\sigma^2}}\label{final:result},
\end{eqnarray}
where the new parameter $\delta$, quantifies the shifting, is defined by
\begin{equation}
\delta=\sqrt[4]{\frac{f(r_A)}{f(r_B)}}-1=\sqrt{\frac{\Omega_B}{\Omega_A}}-1.
\end{equation}
The expression of $\frac{\Omega_B}{\Omega_A}$ in the equatorial plane of the Kerr spacetime has been shown in works \cite{15,16,17}, which is given by
\begin{equation}\label{aw}
\frac{\Omega_B}{\Omega_A}=\frac{1+\epsilon \frac{a}{r_B}\sqrt{\frac{M}{r_B}}}{C\sqrt{1-3\frac{M}{r_B}+
2\epsilon\frac{a}{r_B}\sqrt{\frac{M}{r_B}}}},
\end{equation}
where $C=[1-\frac{2M}{r_A}(1+2a {\omega})+\big(r^2_A+a^2-\frac{2Ma^2}{r_A}\big){\omega}^2]^{-\frac{1}{2}}$ is the normalisation constant, $\omega$ is the Earth's equatorial angular velocity and $\epsilon=\pm1$ stand for the direct of orbits (i.e., when $\epsilon=+1$ for the satellite co-rotates with the Earth).

We notice that $(r_A \omega)^2>a\omega$, therefore we can retain the second order terms in $r_A\omega$. Expanding Eq. (\ref{aw}) we
obtain the following perturbative expression of $\delta$. This perturbative result does not depend
on whether the Earth and the satellite are co-rotating or not.
\begin{eqnarray}\label{bw}
\delta&=&\delta_{Sch}+\delta_{rot}+\delta_h\\
\nonumber&=&\frac{1}{8}\frac{r_S}{r_A}\big(\frac{1-2\frac{h}{r_A}}{1+\frac{h}{r_A}} \big)-\frac{(r_A\omega)^2}{4}-\frac{(r_A\omega)^2}{4}\big(\frac{3}{4}\frac{r_S}{r_A}-\frac{4Ma}{\omega r_A^3}\big),
\end{eqnarray}
where $h=r_B-r_A$ is height between Alice and Bob, $\delta_{Sch}$ is the first order Schwarzschild term, $\delta_{rot}$ is the lowest order rotation term and $\delta_h$ is all the higher order corrections term. Within the geostationary orbits for satellites, we can estimate the order of magnitude of the parameters, which are $\delta_S\sim -10^{-10}$, $\delta_{rot}\sim-10^{-12}$ and $\delta_h\sim-10^{-21}$, respectively. Therefore, we consider $\delta$ as a small quantity which reserves second order
term, the Eq. (\ref{theta}) can be approximate to $\Theta \sim1-\frac{\delta^2\Omega_{B,0}^2}{8\sigma^2}$.

\section{The effect of the Earth's curved space-time on quantum key distribution}
\subsection{The final state affected by the gravitational field of the Earth}
We want to study the QKD protocols described  previously in the framework of photon exchange between the
ground and satellites. The two reliable parties Alice, who is stationary on the Earth's surface and Bob, who is stationary on the satellite, and will receive the photon from Alice.  Assuming that we prepare a two-mode squeezed state in advance (with the modes of $a$ and $b$ at the
ground station), the covariance matrix $AA^{'}$ is given by
\begin{equation}\label{matrix}
\gamma_{AA^{'}}=
\left(
\begin{array}{cc}
V I_2& \sqrt{V^2-1}Z\\
\sqrt{V^2-1}Z&V
\end{array}
\right),
\end{equation}
where $Z=\left(\begin{array}{cc}1& 0\\0&-1\end{array}\right)$. After one mode sends to Bob through the noisy channel caused by Eve (please see Eq. (1)),
$\mathcal{N}_{A^{'}\rightarrow B}(\gamma_{AA^{'}})=\gamma_{AB}$, the covariance matrix of the output can be easily computed
\begin{equation}\label{1}
\gamma_{AB}=
\left(
\begin{array}{cc}
V I_2& \sqrt{T(V^2-1)} Z\\
\sqrt{T(V^2-1)} Z&T( V+\chi) I_2
\end{array}
\right).
\end{equation}

However, Alice sends photon with mode $b$ propagating from the Earth to the satellite and then received by Bob. Due to the Earth's gravitational field, the wave packet of the received photon will be deformed. Similarly, the propagating process can be described by the following form, which has adequate discussion in refs \cite{15,16,17}
\begin{eqnarray}
\bar{b}&=&\Theta\,b+\sqrt{{1-\Theta^2}}b_{\bot}.
\end{eqnarray}

The mixing beam splitting of modes $b$ and $b_{\bot}$ will represent this process. For this process, the symplectic transformation can be encoded into the Bogoloiubov transformation
\begin{equation}
\gamma_{BS}=
\left(
\begin{array}{cc}
\Theta I_2& \sqrt{1-\Theta^2}I_2\\
-\sqrt{1-\Theta^2}I_2&\Theta I_2
\end{array}
\right).
\end{equation}

For the curved spacetime background of the Earth, the covariance matrix of the transmitted state can be obtained by
\begin{eqnarray}
\gamma_{ASBS}=(Y_{BS})^T(\gamma_{AB}\oplus\gamma_\nu)(Y_{BS}),
\end{eqnarray}
where $\gamma_\nu$ is the covariance matrix of the vacuum state and the matrix $Y_{BS}=I_2\oplus\gamma_{BS}$.
We trace over the irrelevant modes, as a result, the covariance matrix $\gamma_{AB}^{final}$ is given by
\begin{equation}
\gamma_{AB}^{fianl}=
\left(
\begin{array}{cc}
V I& \Theta\sqrt{T(V^2-1)} Z\\
\Theta\sqrt{T(V^2-1)} Z&T[(\Theta^2 V+1-\Theta^2)+\chi] I
\end{array}
\right).
\end{equation}
Here, the wave packet overlap parameter $\Theta$ contains the effect of the Earth's gravitational field. We notice again that $T$ is transmissivity of noise channel. In fact, the transmissivity of channel loss contains many complex effects. We give a brief introduction below.
The channel losses contain four main parts effects, i.e., free-space diffraction, atmospheric extinction, setup-loss (due to non-unit quantum efficiency of the detector and other optical imperfections), and fading (turbulence and pointing errors). Assuming Alice (transmitter) stays at the ground station and Bob (receiver) stays at some variable satellite height $h$ with a variable zenith angle $\theta$. This angle cause by the zenith point at the ground station with the direction of observation pointing at the satellite, and free-space quantum communication is based on a quasi-monochromatic optical mode.
The total amount of loss of the free-space channel from the generation of the Gaussian beam
to its final detection is described by following form
\begin{equation}
T_{tot}(h,\theta)=T_{eff}T_{atm}(h,\theta)T_{d}(h,\theta),
\end{equation}
where $T_{eff}$ is setup-loss transmissivity and assumes as $T_{eff}\simeq0.4$ \cite{27a}. $T_{atm}(h,\theta)=\exp^{-\alpha_0g(h,\theta)}$ expresses for the atmospheric transmissivity, where  $\alpha_0\simeq 5\times 10^{-6} m^{-1}$ is the extinction factor at the sea-level $h_0=6600$ m \cite{27b}, and $g(h,\theta)$ is integral function. $T_{d}(h,\theta)=1-\exp^{-\frac{2a_R^2}{w_d[z(h,\theta)]^2}}$ is the diffraction-induced transmissivity, where $w_d$ is increased spot size after travelling a slant distance $z$ by diffraction effect, $a_R$ is the aperture of the receiving telescope, and $z(h,\theta)=\sqrt{h^2+2hr_A+r_A^2\cos^2\theta}-r_A\cos\theta$ with $r_A\simeq6371$ Km. For further details about transmissivity of channel loss, we refer readers to the literature \cite{27,28,29}.
The relation between transmissivity and channel loss is $T=10^{-l/10}$, where $l=0.2$ dB/Km is the channel loss \cite{21a}. The channel loss as a function of the satellite height is  shown in Fig. 1.
In addition, if Eve were to attack, the input signal will become  $\chi=\frac{1-T}{T}+\epsilon$ , where $\epsilon$ is the so-called excess noise, and this noisy channel is caused by Eve.
\begin{figure*}[t!]
	\begin{center} {\includegraphics[scale=0.8]{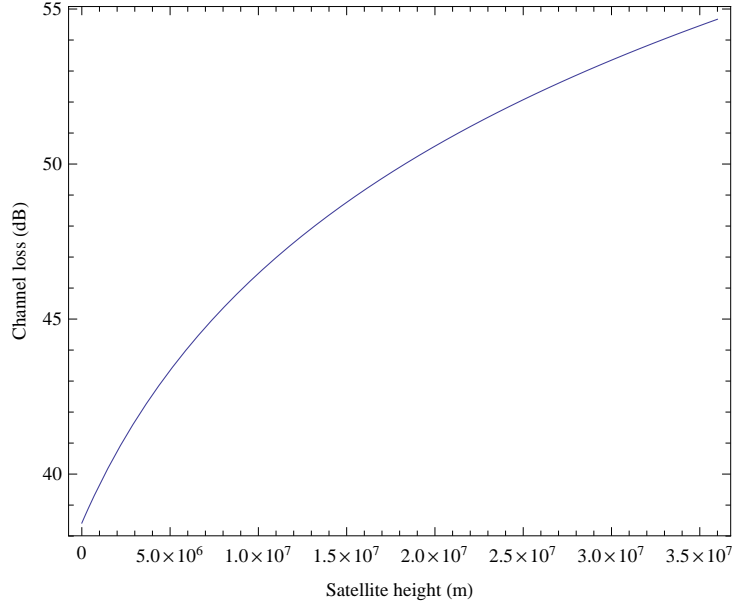}}
		\caption{ The channel loss as a function of the satellite height. \label{fig1.1}}
	\end{center}
\end{figure*}


\subsection{Key rates in the Earth's curved space-time}
According to the context in section 2, one can calculate the mutual information in this case, which is given by following form
\begin{eqnarray}
I(a,b)=\frac{1}{2}\log(\frac{r}{r-t^2/s}),
\end{eqnarray}
where $r=V$, $s=T[(\Theta^2 V+1-\Theta^2)+\epsilon]$ and $t=\Theta\sqrt{T(V^2-1)}$.
The second term in secret key rate $K$,  the Holevo bound for DR can be written as
\begin{eqnarray}
S(a:E)=S(E)-S(E|a)=g(\lambda_1)+g(\lambda_2)-g(\lambda_3),
\end{eqnarray}
where now $\lambda_{1,2}$ are the symplectic eigenvalues of
\begin{eqnarray}
\lambda_{1,2}&=&\sqrt{\frac{1}{2}(\triangle\pm\sqrt{\triangle^2-4D^2})},\nonumber\\
\triangle&=& r^2+s^2-2t^2,\nonumber\\
D&=&rs-t^2,
\end{eqnarray}
and $\lambda_3=s$ is the symplectic eigenvalues of $\sigma^{(d)}_{B|a}$.
Then, the final secret key rate of QKD protocol under the Earth's curved space-time is
\begin{equation}
K=\frac{1}{2}\log(\frac{r}{r-t^2/s})-g(\lambda_1)-g(\lambda_2)+g(\lambda_3).
\end{equation}

\section{Results and discussions}

\begin{figure*}[t!]
	\begin{center} {\includegraphics[scale=0.85]{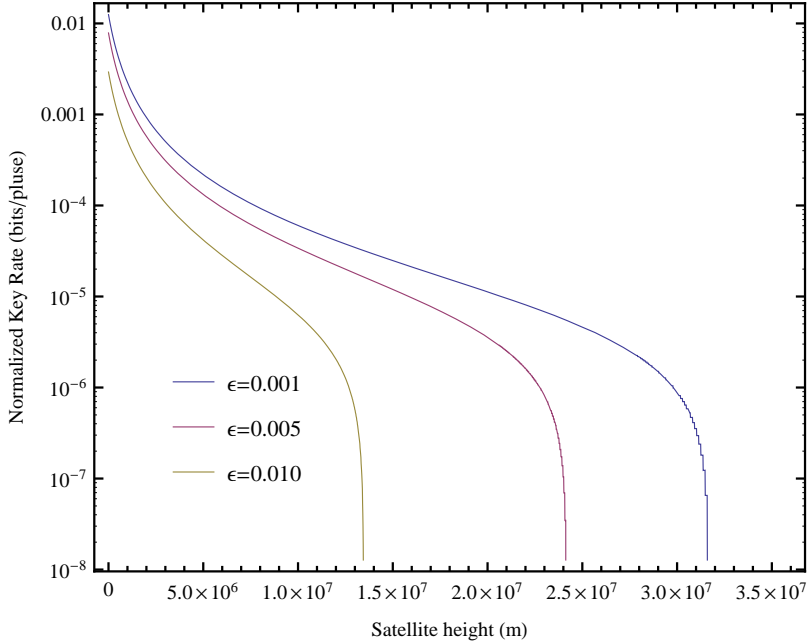}}
		\caption{The normalized key rates $K$ as the functions of satellite height with the different excess noises $\epsilon=0.001$ (blue), $\epsilon=0.005$ (purple), $\epsilon=0.010$ (yellow).  The other parameters are fixed as $V=2$ and $\Omega_0=500$THz.}   \label{fig1}
	\end{center}
\end{figure*}
It should be noted that the two terminals of the quantum key distribution task are only a few hundred kilometers apart based on the current technology. The effect of gravity on the quantum key rate depends on the gravitational potential difference, i.e., the height difference. Therefore, our work only analyzes the influence of gravitational effects on future satellite-based long-distance quantum key missions within geostationary Earth orbits from a theoretical perspective.  In other words, this is a study of pure gravitational effects that have nothing to do with the acceleration of the satellite itself, for example, the Unruh effect of a massive, accelerating object. Our work focuses on the effects of the Earth's space-time background on the quantum secret key, including special relativity effect, but it is not a major part of our analysis.

In Fig.~\ref{fig1}, we plot key rates $K$ as the functions of satellite height $h$ (within geostationary satellite orbit) with a fixed variance of the initial shared entangled state $V=2$ for different excess noises $\epsilon=0.001$, $\epsilon=0.005$, and $\epsilon=0.010$ under the gravitational effect of the Earth. The relation between satellite height $h$ and channel loss $l$ is shown in Fig. 1.  Recently, the orbital angular momentum (OAM) beam has attracted a lot of attention because of its unique optical characteristics, that is, it has the Laguerre-Gaussian amplitude distribution \cite{1992PhRvA..45.8185A}. Due to characteristic of OAM beam carrying orbital angular momentum, it can be applied to particle manipulation \cite{2011NaPho...5..343P}, quantum information \cite{2014NaPho...8..234N}, and other fields \cite{2012NaPho...6..488W}. In our work, we adopt the assumption of Gaussian beams, and the initial Gaussian wave packet parameters, i.e., the peak frequency $\Omega_{0}=500$ THz, the Gaussian bandwidth $\sigma=1$ MHz are considered the taking values of these parameters come from a typical parametric down converter crystal source with a wavelength of 598 nm \cite{24,25}.  We can see that a monotonic decrease of $K$ with the increasing of the channel loss length $l$. Compared with the channel loss, the effects of excess noises on the key rate is almost negligible. Furthermore, the key rates of all the protocols become practically zero at sufficiently high channel loss.  Our results are agreement with the previous works \cite{22,26}, which demonstrates the validity of our method and results. However, it is worth noting that the attenuation factors of quantum secret key include two parts, the first part is the loss of transmission channel, the second part is the influence of Earth's gravitational field.

\begin{figure*}[t!]
	\begin{center} {\includegraphics[width=8cm,height=6.5cm]{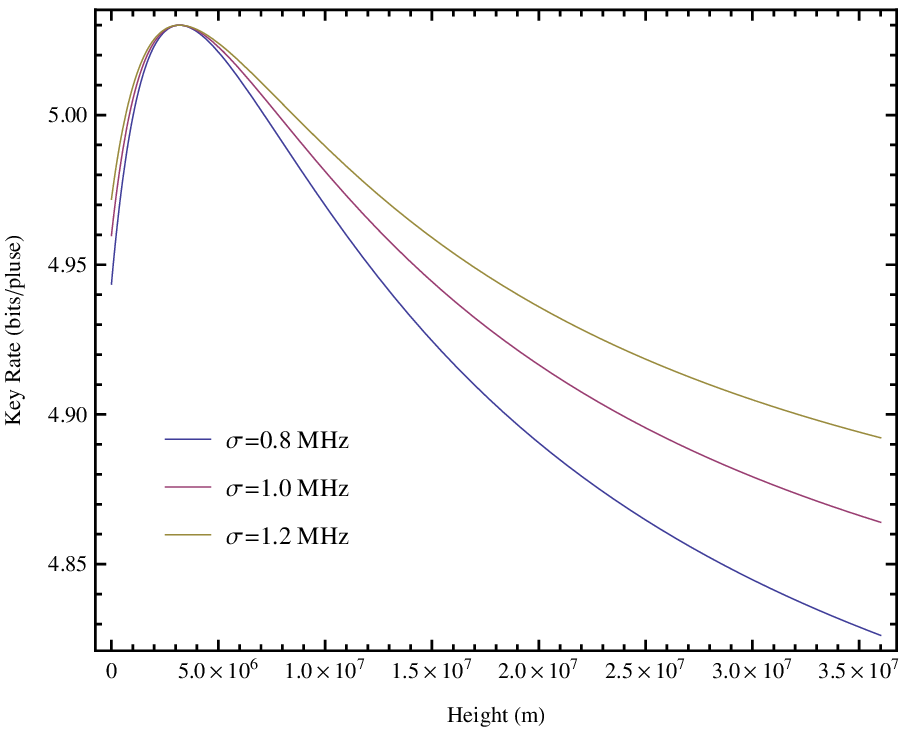}
\includegraphics[width=8cm,height=6.6cm]{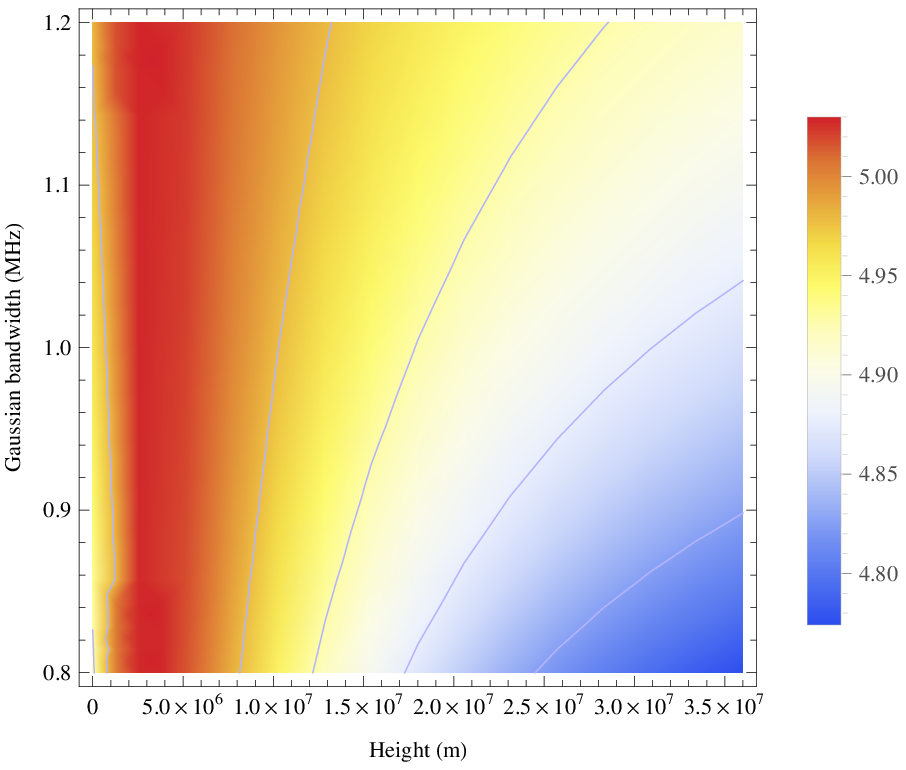}}
		\caption{Left: The key rates $K$ as the functions of orbit heights of satellite with the different Gaussian bandwidth $\sigma=0.8$ MHz (blue), $\sigma=1.0$ MHz  (purple), $\sigma=1.2$ MHz (yellow). The other parameters are fixed as $V=20$, $\epsilon=0.001$ and $\Omega_0=500$THz. Right: Density plot for key rate $K$.   \label{fig2}}
	\end{center}
\end{figure*}
In order to show the influence of the gravitational effect of the Earth on QKD more intuitively, we plot key rates $K$ as the functions of the orbits of satellite $h$ in the left panel of Fig.~\ref{fig2}. We fix the $V=10$, $\epsilon=0.001$ and the peak frequency $\Omega_{0}=500$ THz.  Not surprisingly, the Earth's gravitational field erodes the ability of the two parties to perform QKD in all the protocols. It also shows some interesting features: The key rates initially increase for a specific range of height parameter $h\simeq r_A/2$, and then gradually decrease with the increasing of the satellite orbits $h$. The possible explanation we give is based on the fact that gravitational frequency shift and special relativistic effects play different roles in key rates. When the reference party Bob (satellite) stays at the height $h=r_A/2$ (half the radius of the Earth) accompanied with vanishing Schwarzschild term $\delta_{Sch}=0$,  the photons received on satellites will suffer Doppler blue shift effect ($\delta>0$). In the case, the lowest-order rotation term $\delta_{rot}$ and higher-order correction term $\delta_h$ should be taken into consideration, which causes the Doppler blue shift effect. That's why we didn't use classical Newtonian gravity to consider our case.  As the altitude of the satellite increases, the special relativistic effects become smaller while the gravitational frequency shift can be cumulated. On the other hand, one can see that the initial Gaussian bandwidth parameter has a great influence on the key rate. This result can be clearly seen in the density plot on the right panel of Fig.~\ref{fig2}. This gives us the guidance to choose appropriate physical parameters to perform more reliable QKD with the inevitable relativistic effects.

\begin{figure*}[t!]
	\begin{center} {\includegraphics[scale=0.85]{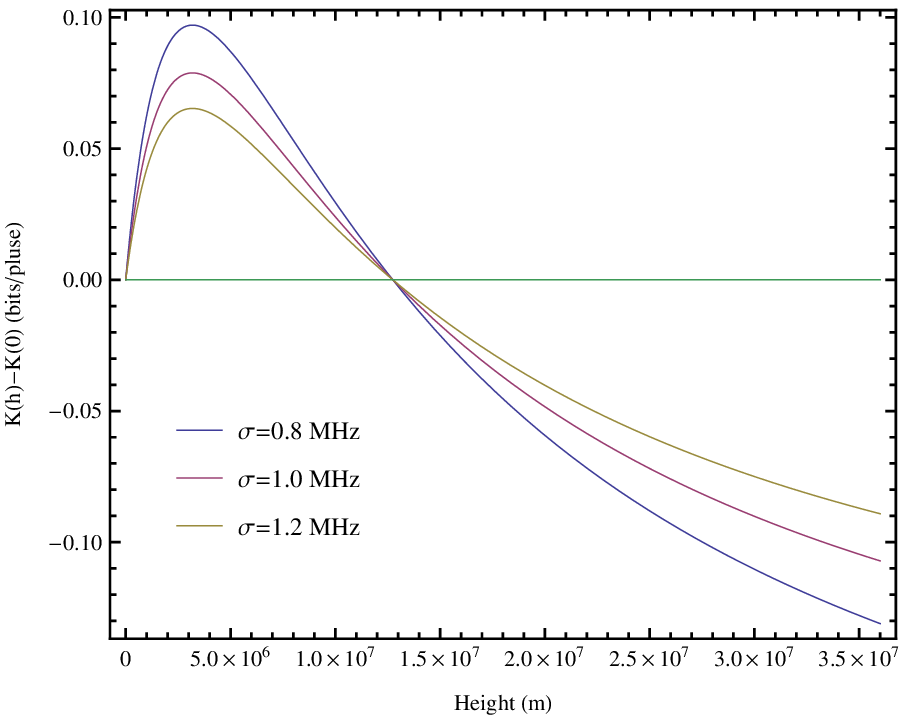}
\includegraphics[scale=0.875]{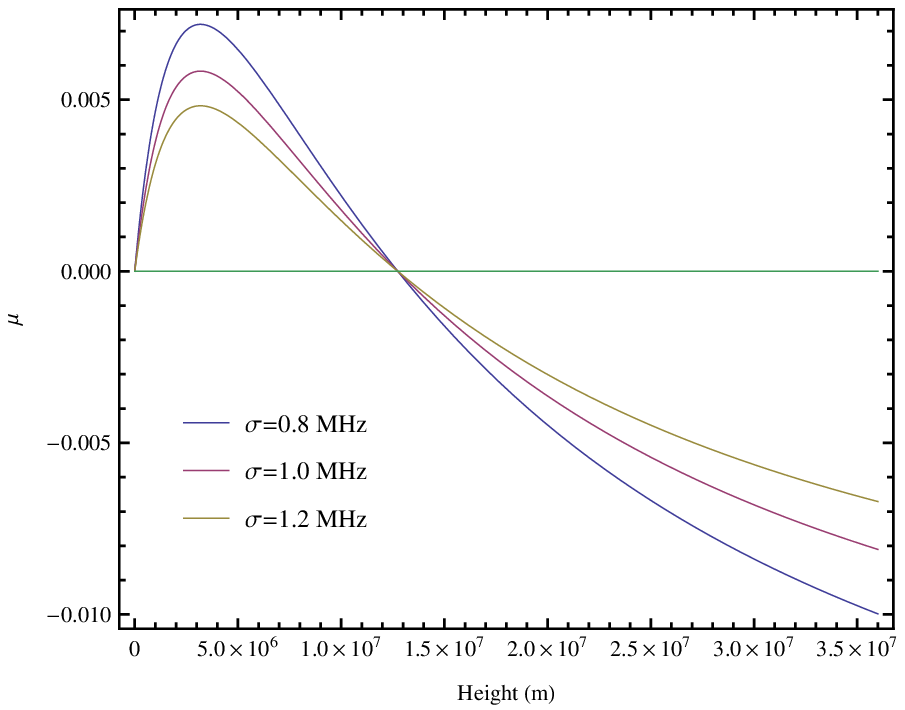}}
		\caption{Left: The difference of key rates $K$ with the variations of orbit heights of satellite $h$  with the different the initial Gaussian bandwidth parameters $\sigma=0.8, 1.0, 1.2$ MHz. Right: Same as on the left, but for change rate of key rate $\mu$. \label{fig3}}
	\end{center}
\end{figure*}

To better demonstrate the effect of Earth's gravity on the quantum key rate, we firstly plot the key rate difference $K(h)-K(0)$ in the left panel of Fig~\ref{fig3}, with the different the initial Gaussian bandwidth parameters $\sigma=0.8, 1.0, 1.2$ MHz. The other parameters are fixed $V=10$, $\epsilon=0.001$, and $\Omega_{0}=500$ THz. The $K(0)$ represents the key rate without Earth's gravity, which corresponds value at the satellite orbit height $h=0$ and corresponds a perfect channel $\Theta=1$. With the aim of furthermore quantifying the influence of the Earth's curved space-time, we will define an additional quantity to describe the change rate of the key rate
\begin{equation}
\mu=\frac{K(h)-K(0)}{K(0)}.
\end{equation}
In the right panel of Fig~\ref{fig3}, we plot the change rate of key rate  $\mu$ as a function of
the height parameter $h$. One can clearly see that the change in key rate generated by gravitational frequency
shift can be determined at a level of $<1.0\%$ within the satellite height at geostationary Earth orbits.   Such findings could provide some interesting possibilities to reduce the loss key rate, through the control of the orbital height of satellites. The Earth's gravitational field does not have a great influence on the secret key rate, because the Earth's gravitational field is not strong enough. However, according to the equivalence principle of general relativity, our results can be applied to the accelerated coordinate system equivalently, and we can infer that when the acceleration is large enough, relativistic effects will significantly affect the key rate.

\section{Conclusion}

In this work, we have shown a very general method to consider the effects of the Earth's gravitational field on the CV QKD protocols. In realistic situation, the preparation of quantum system and the procession of quantum information tasks are always accompanied with gravitational and relativistic effects. Our work highlights the importance of studying the CV QKD protocols in the space-time background of the Earth.

Not surprisingly, the Earth's gravitational field erodes the ability of the two parties to perform QKD in all the protocols. However, our findings also have some interesting features. The key rates initially increase for a specific range of height parameter $h\simeq r_A/2$, and then gradually decrease with the increasing of the satellite orbits $h$. The possible explanation we give is based on the fact that gravitational frequency shift and special relativistic effects play different roles in key rates. When the reference party Bob (satellite) stays at the height $h=r_A/2$ (half the radius of the Earth) accompanied with vanishing Schwarzschild term $\delta_{Sch}=0$,  the photons received on satellites will have doppler blue shift effect ($\delta>0$). Further, we demonstrate that the key rates are  monotonic decreasing with the increasing of the channel losses lengths $l$. It should be noted the monotonic decrease of $K$ with the increasing of the channel loss length $l$. Compared with the channel loss, the effects of excess noises on the key rate are almost negligible.  This results are agreement with the previous works \cite{22,26}, which demonstrates the validity of our method and results.  Moreover, with the increase in the variance of the initial shared entangled state, the key rate has significantly increased. This gives us the guidance to choose appropriate physical parameters to perform more reliable QKD with the inevitable relativistic effects.
To better demonstrate the effect of Earth's gravity on the quantum key, we show that the change of key rate generated by gravitational frequency shift can be determined at the level of $<1.0\%$ within the satellite height at geostationary Earth orbits.   Such findings could provide some interesting possibilities to reduce the loss key rate, through the control of the orbital height of satellites.

With the rapid development of quantum technology, quantum communication and quantum information tasks have recently expanded to include satellites orbiting the Earth. Long distance communication protocols cannot ignore the existence of gravitational field and its effect on quantum states. Therefore, our work is of fundamental and practical significance. Meanwhile, according to the equivalence principle of general relativity, our results can be applied to the accelerated coordinate system equivalently.
On the other hand, the Gaussian modulations in the QKD protocols have the obvious advantage, because its associated homodyne or heterodyne detection scheme offers the prospect of very high key rates, where homodyne detection measures one quadrature of the field, and heterodyne detection provides a joint measurement of both quadrature and phase. However, the discrete modulation reflects what is actually done in experiments. For more on discrete modulation work of QKD protocols we refer the reader to the literature \cite{30,31,32}.  If we consider the Earth's gravitational field as a loss channel as in this work, our model can in principle be applied to  key distribution protocols of discrete modulation with coherent state and coherent detection, thus providing a theoretical basis for future quantum key distribution techniques to some extent.

\textbf{Data availability}
The datasets analysed during the current study are available from the corresponding author
on reasonable request.

\textbf{Competing Interests Statement}
The authors have no competing interests to declare that are relevant to the content of this article.

\textbf{Acknowledgements} We are grateful to our anonymous reviewers for their advice on our work, which has greatly improved the quality of our article. This work was supported by the National Natural Science Foundation of China under Grants Nos. 12203009, 12065016, 12021003, and 11690023; the Strategic Priority Research Program of the Chinese Academy of Sciences, Grant No. XDB23000000; the Young scientific talents growth project of the department of education of the department of education of Guizhou province under Grant No. QJHKYZ[2019]129.


\begin{thebibliography}{}


\bibitem{1}Bennet C H, and  Brassard G, in Proceedings IEEE International Conference on Computers, Systems and
Signal Processing, Bangalore, India (IEEE, New York, 1984), p. 175

\bibitem{1a}Ekert A K 1991 \textit{Phys. Rev. Lett.} {\bf67} 661 doi:10.1103/PhysRevLett.67.661

\bibitem{2}Cerf  N  J and  Grangier P 2007 \textit{J. Opt. Soc. Am. B}  {\bf24} 324

\bibitem{2a} Grosshans F,   Van Assche G,  Wenger J,  Brouri R,
Cerf N J and Grangier P 2003 \textit{Nature} {\bf421} 238

\bibitem{2b}Aspelmeyer M, B{\"o}hm H R, Gyatso T et al. 2003 \textit{Science} {\bf301} 621

\bibitem{2c}Dynes J F, Takesue H, Yuan Z L et al. 2009  \textit{Optics Express} {\bf17} 11440

\bibitem{3}Yin J,  Ren J G,  Lu H et al. 2012 \textit{Nature} {\bf488} 185

\bibitem{4}Yin J. et al. 2017 \textit{Science} {\bf356} 1140

\bibitem{5}Nadlinger D,  Drmota P, Nichol B et al. 2021 \textit{ arXiv preprint} arXiv:2109.14600

\bibitem{6}Zhang W,  Leent T van,  Redeker K,  Garthoff R,
Schwonnek R,  Fertig F, Eppelt  S,  Scarani V, Lim C C
W and  Weinfurter H 2021 \textit{arXiv preprint} arXiv:2110.00575

\bibitem{7}Liu  W Z,  Zhang Y Z, Zhen  Y Z,  Li M H,  Liu Y,
 Fan J, Xu F, Zhang Q and Pan  J W 2021 \textit{ arXiv preprint} arXiv:2110.01480


\bibitem{8}Lo H K, Curty M and  Qi B 2012 \textit{ Phys. Rev. Lett.}
{\bf108} 130503

\bibitem{9}Lucamarini M,  Yuan Z L, Dynes J F and
Shields A J 2018 \textit{ Nature} {\bf557}
400

\bibitem{9a}Xu F,  Ma X,  Zhang Q, Lo H K and Pan J W 2020 \textit{Rev. Mod. Phys.} {\bf92} 025002

\bibitem{9b}Pirandola S,  Andersen U L,  Banchi L et al. 2020 \textit{Adv. Opt. Photonics} {\bf12} 1012

\bibitem{9c}Ac\'{\i}n A, Brunner  N,  Gisin N, Massar S, Pironio  S
and  Scarani V 2007  \textit{Phys. Rev. Lett.}
{\bf98} 230501

\bibitem{21a} Singh J and Bose S  2021 \emph{Phys. Rev. A} {\bf104} 052605


\bibitem{10b}Braunstein S L, and  Pirandola S 2012
\textit{Phys. Rev. Lett.} {\bf108} 130502

\bibitem{10q} Pirandola S,  Laurenza R,  Ottaviani C and  Banchi L 2017
\textit{Nature communications} {\bf8} 15043

\bibitem{10c}Ma X C, Sun S H, Jiang  M S, Gui M and
Liang L M 2014 \textit{Phys. Rev. A} {\bf 89} 042335

\bibitem{10d}  Pirandola S,  Ottaviani C,  Spedalieri G et al. 2015 \textit{Nature Photonics} {\bf 9} 397

\bibitem{10e}  Zhang Y C, Li  Z,  Yu S,  Gu W,  Peng X and  Guo H 2014
\textit{Phys. Rev. A} {\bf 90} 052325

\bibitem{10f} Papanastasiou P,  Ottaviani C and  Pirandola S 2017
\textit{Phys. Rev. A} {\bf 96} 042332

\bibitem{10g}  Gehring T,  \"{H}andchen V,  Duhme J,  Furrer F,
 Franz T,  Pacher C, Werner R F and Schnabel R 2015
\textit{Nature Communications} {\bf6} 8795

\bibitem{10h} Xu F,  Curty M,  Qi B, Qian  L and  Lo H K 2015
\textit{Nature Photonics} {\bf9} 772

\bibitem{10}Scarani V,  Bechmann-Pasquinucci H, Cerf N J,
Du\v{s}ek M, L\"{u}tkenhaus N and  Peev M 2009 \textit{Rev. Mod. Phys.}
{\bf81} 1301

\bibitem{11}Weedbrook C, Pirandola S, Garc\'{\i}a-Patr\'{o}n  R, Cerf N J,
 Ralph T C,  Shapiro J H and Lloyd S 2012 \textit{Rev. Mod. Phys}.
{\bf84} 621


\bibitem{CVQKB1} Soh D B~S, Brif C, Coles P~J et al.\ 2015 \textit{Physical Review X} {\bf5} 041010

\bibitem{CVQKB2} Qi B, Lougovski P, Pooser R et al.\ 2015 \textit{Physical Review X} {\bf5} 041009

\bibitem{CVQKB3} Jouguet P, Kunz-Jacques S, Leverrier A et al.\ 2013 \textit{Nature Photonics} {\bf7} 378

\bibitem{CVQKB4} Zhou Y-H, Qin S-F, Shi W-M et al.\ 2022 \textit{Quantum Information Processing} {\bf21} 303

\bibitem{CVQKB5} Zhao W, Shi R, Ruan X et al.\ 2022 \textit{Quantum Information Processing} {\bf21} 186
\bibitem{CVQKB6} Aguiar L S, Borelli L~F~M, Roversi J~A et al.\ 2022 \textit{Quantum Information Processing} {\bf21} 304
\bibitem{CVQKB7} Peng Q, Guo Y, Liao Q et al.\ 2022 \textit{Quantum Information Processing} {\bf21} 61
\bibitem{CVQKB8} Zhang M, Dou Y, Huang Y et al.\ 2021 \textit{Quantum Information Processing} {\bf20} 327
\bibitem{CVQKB9} Liu B, Xia S, Xiao D et al.\ 2022 \textit{Science China Physics, Mechanics, and Astronomy} {\bf65} 240312
\bibitem{CVQKB10} Zhou C, Wang X, Zhang Z et al.\ 2021 \textit{Science China Physics, Mechanics, and Astronomy} {\bf64} 260311
\bibitem{CVQKB11} Xie Y-M, Lu Y-S, Weng C-X et al.\ 2022 \textit{PRX Quantum} {\bf3} 020315
\bibitem{CVQKB12} Zeng P, Zhou H, Wu W, et al.\ 2022 \textit{Nature Communications} {\bf13} 3903
\bibitem{12}Grochowski  P T, Rajchel G, Kialka  F and  Dragan A 2017
\textit{Phys. Rev. D} {\bf95} 105005

\bibitem{13}Ahmadi  M,  Lorek K,  Ch\c{e}ci\'{n}ska A,  Smith A R H,
Mann R B and  Dragan A 2016 \textit{Phys. Rev. D} {\bf93} 124031

\bibitem{14}Richter B,  Lorek K,  Dragan A and  Omar Y 2017 \textit{Phys. Rev. D}
{\bf95} 076004

\bibitem{wu1}Wu S M and Zeng H S 2022  \textit{Eur. Phys. J. C} {\bf82} 716

\bibitem{wu2}Wu S M, Zeng H S and Liu T H 2022 \textit{New J. Phys.} {\bf24} 073004

\bibitem{wu3}Wu S M, Cai Y T, Peng W J and Zeng H S 2022 \textit{Eur. Phys. J. C}? {\bf82} 412

\bibitem{Visser}
Visser M  2007 arXiv:0706.0622

\bibitem{kerr}
Kohlrus J,  Bruschi D,  Louko J  and Fuentes I 2017
\textit{EPJ Quantum Technology }{\bf4}, 7


\bibitem{ULMQ}
 Leonhardt U \emph{Measuring the Quantum State of Light,
Cambridge Studies in Modern Optics} (Cambridge University
Press, Cambridge, 2005).


\bibitem{TGDT}
Downes T,  Ralph T and  Walk N 2013 \emph{Phys. Rev. A}
{\bf 87} 012327

\bibitem{DEBT}
 Bruschi D,  Ralph T,  Fuentes I,  Jennewein T
and  Razavi M  2014 \emph{Phys. Rev. D } {\bf 90} 045041

\bibitem{DEBA}
 Bruschi D,  Datta A, Ursin  R,  Ralph T and  Fuentes I  2014
\emph{Phys. Rev. D}  {\bf90} 124001

\bibitem{PPRW}
 Rohde P,  Mauerer W and Silberhorn  C 2007 \emph{New Journal
of Physics} {\bf 9 } 91

\bibitem{15}
Liu T H,  Cao S and Wu S M 2020 \textit{Sci. Rep.} {\bf10} 14697

\bibitem{16}
Liu T H,  Cao S,  Wu S M and  Zeng H S 2019 \textit{Laser Phys. Lett.} {\bf 16} 095201

\bibitem{17}
 Liu T H, Jing J L and  Wang J C 2018 \textit{Adv. Quantum Technol.} {\bf1} 1800072

 \bibitem{18}Liu T H, Wang J C and Jing J L et al. 2018 \textit{Annal of Physics} {\bf390} 334

\bibitem{19}Fang Y, Liu X, Wang J et al. 2019 \textit{Quantum Information Processing} {\bf18} 248

\bibitem{20} Zhang S X, Liu T H, Cao S et al. 2020 \textit{Chinese Physics B} {\bf29} 050402.

\bibitem{21} Gong X L, Cao S, Fang Y, and Liu T H 2022 \textit{Chinese Physics B} {\bf31} 050402

\bibitem{22} Pierini R 2018 \textit{Phys. Rev. D} {\bf98} 125007
\bibitem{23} Weedbrook C, Pirandola S, Garciapatron R et al. 2012 \textit{Rev. Mod. Phys.}
{\bf84} 621
\bibitem{24} Razavi M and Shapiro J 2006 \textit{Phys. Rev. A} {\bf73} 042303
\bibitem{27a}Vasylyev D,  Vogel W and  Moll F 2019 \textit{Phys. Rev. A} {\bf 99} 053830

\bibitem{27b}  Liorni C,  Kampermann H and  Bru{\ss} D 2019 \textit{New J. Phys.}  {\bf 21} 093055
\bibitem{27} Dequal D, Trigo Vidarte L, Roman Rodriguez V et al.\ 2021 \textit{npj Quantum Information} {\bf7} 3
\bibitem{28} Pirandola S\ 2021 \textit{Physical Review Research} {\bf3} 013279
\bibitem{29} Pirandola S\ 2021 \textit{Physical Review Research} {\bf3} 023130

\bibitem{25} Matsukevich D, Maunz P, Moehring D, Olmschenk
S and Monroe C 2008 \textit{Phys. Rev. Lett}. {\bf100} 150404
\bibitem{1992PhRvA..45.8185A}
 Allen L, Beijersbergen M~W, Spreeuw R~J~C et al. 1992 \textit{Phys. Rev. A} {\bf45} 8185
\bibitem{2011NaPho...5..343P}Padgett M and Bowman R 2011 \textit{Nature Photonics} {\bf5} 343
\bibitem{2014NaPho...8..234N}Nicolas A, Veissier L, Giner L et al. 2014 \textit{Nature Photonics} {\bf8} 234
\bibitem{2012NaPho...6..488W}Wang J, Yang J Y, Fazal I~M et al.\ 2012 \textit{Nature Photonics} {\bf6} 488
\bibitem{26}Garc{\'\i}a-Patr{\'o}n R and Cerf N~J 2009 \textit{Phys. Rev. Lett}. {\bf102} 130501
\bibitem{30}Liu W-B, Li C-L, Xie Y-M et al. 2021  \textit{PRX Quantum} {\bf2} 040334
\bibitem{31}Zhou M-G, Liu Z-P, Liu W-B et al.\ 2022  \textit{Scientific Reports} {\bf12} 8879
\bibitem{32}Liu Z-P, Zhou M-G, Liu W-B et al.\ 2022 \textit{ Optics Express} {\bf30} 15024
\end{thebibliography}

\end{CJK*}
\end{document}